\documentclass[runningheads]{llncs}
\usepackage[T1]{fontenc}
\usepackage{graphicx}
\usepackage{color,xcolor}
\usepackage{multirow}
\usepackage{booktabs}
\usepackage{amsmath}

\usepackage{amssymb}
\usepackage{url}
\usepackage{algpseudocode}
\usepackage{amsfonts}
\usepackage[ruled,linesnumbered]{algorithm2e}
\usepackage{multicol}     
\usepackage{multirow}
\usepackage[misc]{ifsym}
\usepackage{color}

\usepackage{floatrow}
\newfloatcommand{capbtabbox}{table}[][\FBwidth]

\begin{document}
%
\title{Dual-channel Early Warning Framework for \\Ethereum Ponzi Schemes}


%

%
\author{
        Jie Jin\inst{1,2} \and
        Jiajun Zhou\inst{1,2} \textsuperscript{(\Letter)} \and 
        Chengxiang Jin\inst{1,2} \and  \\
        Shanqing Yu\inst{1,2}   \and
        Ziwan Zheng\inst{5}  \and
        Qi Xuan\inst{1,2,3,4}
}
    
\institute{
    Institute of Cyberspace Security, Zhejiang University of Technology, \\Hangzhou 310023, China \\ \and
    College of Information Engineering, Zhejiang University of Technology, \\Hangzhou 310023, China\\ \and
    The PCL Research Center of Networks and Communications, \\ Peng Cheng Laboratory, Shenzhen 518000, China\\ \and
    The Utron Technology Company Limited, Hangzhou 310056, China\\ \and 
    Zhejiang Police College, Hangzhou 310053, China \\
    \email{jjzhou@zjut.edu.cn}   
}
\authorrunning{J. Jin et al.}
%
\maketitle             

\begin{abstract}
Blockchain technology supports the generation and record of transactions, and maintains the fairness and openness of the cryptocurrency system. 
However, many fraudsters utilize smart contracts to create fraudulent Ponzi schemes for profiting on Ethereum, which seriously affects financial security.
Most existing Ponzi scheme detection techniques suffer from two major restricted problems: the lack of motivation for temporal early warning and failure to fuse multi-source information finally cause the lagging and unsatisfactory performance of Ethereum Ponzi scheme detection.
In this paper, we propose a dual-channel early warning framework for Ethereum Ponzi schemes, named \emph{Ponzi-Warning}, which performs feature extraction and fusion on both code and transaction levels.
Moreover, we represent a temporal evolution augmentation strategy for generating transaction graph sequences, which can effectively increase the data scale and introduce temporal information.
Comprehensive experiments on our Ponzi scheme datasets demonstrate the effectiveness and timeliness of our framework for detecting the Ponzi contract accounts.


\end{abstract}

\section{Introduction}
Ponzi scheme~\cite{frankel2012ponzi} is a traditional financial fraud that lures investors into investing by promising high returns, and pays profits to earlier investors with funds from more recent investors, convincing them that the profits came from legitimate business activity.
Traditional Ponzi schemes generally share similar characteristics that deserve high vigilance of the investors: 1) Suspiciously high investment returns with little or no risk; 2) Overly consistent returns; 3) Secretive or complex strategies~\cite{springer2020politics}.
However, with the rapid development of Blockchain technique and its widespread adoption in the field of digital cryptocurrencies, these traditional Ponzi fraud has also infected the digital finance world.
Fraudsters combine Ponzi schemes with blockchain technology (i.e, Smart Contracts) to create a new form of fraud --- the smart Ponzi scheme. 
In Ethereum, the fraudsters create and deploy fraudulent Ponzi contracts, and advertise them as high-return investments, finally swindle money without offline publicity by spreading Ponzi contract addresses to victims in any way such as emails, chat groups, web links, apps, etc.
Although the Ponzi contract code with complex logic is incomprehensible to investors, it is still trusted because of the openness, transparency and immutability characteristics of blockchain technique.
According to a report published by the Cryptoanalysis, a cryptocurrency investigation and risk analysis company, Ponzi schemes, fraudulent ICOs and other forms of fraud are on the rise, causing at least \$725 million in losses so far.
As a result, financial security has become a top priority in the blockchain ecosystem~\cite{morris2017rise}.

Existing Ponzi scheme detection methods in Ethereum mainly concentrate on manual feature engineering and graph representation learning.
The former mainly combines traditional machine learning methods and manual features (i.e. statistical and structural features) of accounts to detect Ponzi contracts. The statistical features can be code-level~\cite{bartoletti2020dissecting,lou2020ponzi} or transaction-level~\cite{jung2019data} while the structural features can be behavior patterns~\cite{zhou2022behavior,jin2022heterogeneous}.
The latter generally constructs Ethereum transaction graphs and utilizes graph representation learning techniques to capture implicit features.
The transaction graphs generally consist of account nodes and transaction edges, while the graph representation learning methods can be random walk~\cite{perozzi2014deepwalk,grover2016node2vec} and graph neural network~\cite{kipf2016semi}.
However, there are still two major restricted problems on existing research of Ponzi detection:
\begin{itemize}
  \item[$\bullet$] \textbf{Lack of motivation for temporal early warning.} The existing methods utilize the last transaction records to detect Ponzi contracts from the moment the fraud has occurred, i.e., following a hindsight perspective, instead of taking temporal information of account behaviors into consideration, which results in failure to detect the occurrence of Ponzi schemes and impose sanctions in a timely manner.
  \item[$\bullet$] \textbf{Failure to fuse multi-source information.} Multi-source information such as contract codes and transaction records is available for Ponzi detection. However, existing methods usually use a single type of features, or simply concatenate multi-source features for Ponzi detection, without adaptive processing and organic fusion.
\end{itemize}

To address the above problems, in this paper, we propose Dual-channel Early Warning Framework (\emph{Ponzi-Warning}) for detecting the Ethereum Ponzi scheme.
We first collect and collate all current known Ethereum Ponzi scheme data involving contract addresses and corresponding labels from different blockchain platforms, yielding an Ethereum Ponzi Scheme Dataset.
Then we propose a Temporal Evaluation Augmentation strategy for generating transaction graph sequences that reserve temporal transaction information of contract accounts.
Finally, we design a dual-channel early warning framework that fuses multi-source information and consists of a code-aware channel and transaction-aware channel.
The code-aware channel can extract contract opcode features via an MLP model, and the transaction-aware channel can capture the structural transaction behavior features via arbitrary GNN models.
The output of the two channels will be concatenated and fed into the classifiers to identify Ponzi contracts.
Extensive experiments are conducted on real-world datasets to verify the effectiveness of \emph{Ponzi-Warning}.

This work has the following contributions:
\begin{itemize}
  \item[$\bullet$] We collect and collate all current known Ethereum Ponzi scheme data involving contract addresses and corresponding labels from different blockchain platforms, yielding an Ethereum Ponzi Scheme Dataset.
  \item[$\bullet$] We propose a temporal evolution augmentation (\emph{TEAug}) strategy for generating transaction graphs of different scales, which can alleviate the data scarcity and imbalance to some extent without using fake data generation techniques.
  \item[$\bullet$] We propose a dual-channel early warning framework (\emph{Ponzi-Warning}) for Ponzi schemes, which can extract and fuse code-level and transaction-level features from raw data, further achieving powerful and timely early warning for Ponzi schemes on Ethereum.
\end{itemize}
The rest of the paper is organized as follows. 
The related work of Ponzi scheme detection is presented in Sec.~\ref{sec:2}. 
The proposed \textit{Ponzi-Warning} framework is detailedly described in Sec.~\ref{sec:3}. 
The experimental results and discussion are presented in Sec.~\ref{sec:4}. 
We conclude our work in Sec.~\ref{sec:5}.

\section{Related Work}
\label{sec:2}
Ponzi schemes in Ethereum not only retain the characteristics of traditional Ponzi schemes, but also make use of smart contracts.

For a newly created contract, there is no transaction record related to it on the blockchain platforms, so we can only judge its legitimacy by using the characteristics of contract codes, that is, to detect whether it is a Ponzi contract.
Bartoletti et al.~\cite{bartoletti2020dissecting} pioneered an Ethereum Ponzi scheme detection method, where they first classified Ponzi schemes into four categories (i.e. tree, chain, waterfall and permission transfer) based on the logic of the contract source code, and further detect Ponzi schemes by using Levenshtein distance to measure the similarity between bytecodes.
Lou et al.~\cite{lou2020ponzi} converted bytecodes into single-channel images and used a convolutional neural network for image recognition to achieve Ponzi scheme detection.

Once the contract is deployed, the transaction record related to it will be generated once the contract is called. 
In this case, the transaction features can also serve to detect whether it is a Ponzi contract.
Jung et al.~\cite{jung2019data} extracted transaction features from the transaction records of different time periods and combined with the opcode features, finally inputting them to different machine learning classifiers for Ponzi detection. 
Hu et al.~\cite{hu2021transaction} used the transaction records to analyze the investment and return features, and trained LSTM models with the contract transaction data for future Ponzi scheme detection. 

In the final stage of a Ponzi scheme, the complete transaction history is available.
At this point, most of the work is to combine the complete transaction features with the code features for Ponzi scheme detection.
Chen et al.~\cite{chen2018detecting,chen2019exploiting} used machine learning methods to identify Ponzi contracts by analyzing account features and opcode features. 
Zhang et al.~\cite{zhang2021detecting} considered bytecode features on top of existing features, again detected by the LightGBM model. 

\section{Methodology}
\label{sec:3}

To be clear, the Ponzi scheme studied in this paper is a blockchain financial fraud generated by the Ponzi contract deployed on the Ethereum platform to attract external investment, excluding the Ponzi scheme that uses email, social media and other means to defraud under the guise of spreading the concept of ``blockchain''.

  \begin{table*}
    \begin{floatrow}
    \capbtabbox{
      \tiny
      \resizebox{0.45\textwidth}{!}{%
      \begin{tabular}{ccccccc} 
        \hline\hline
        \multirow{2}{*}{\begin{tabular}[c]{@{}c@{}}Ponzi \\Address\end{tabular}} & \multirow{2}{*}{Num.} & \multicolumn{2}{c}{Feature} & \multicolumn{3}{c}{Transaction Num.}  \\ 
        \cline{3-7}
                  &   & \multicolumn{1}{c}{code} & \multicolumn{1}{c}{trans.}  & \multicolumn{1}{c}{Min.} & \multicolumn{1}{c}{Max.}  & \multicolumn{1}{c}{$\geqslant$ 100} \\  
        \hline
        code(only)                            & 68                    & $\surd$ &                         & 0         & 0       & 0                       \\
        code \& trans.                        & 230                   & $\surd$ & $\surd$                 & 1         & 105005 & \textbf{75}           \\ 
        \hline
        all                                   & 298                   &         &                         & 1         & 105005 & \textbf{75}           \\ 
        \hline\hline
        \end{tabular}}
    }{
      \caption{Statastics of the Ethereum Ponzi contract addresses we collected.}
      \label{tab: ponzi}
    }
    \capbtabbox{
      \tiny
      \resizebox{0.45\textwidth}{!}{%
      \begin{tabular}{ccccccc} 
        \hline\hline
        \multirow{2}{*}{\begin{tabular}[c]{@{}c@{}}Contract \\Address\end{tabular}} & \multirow{2}{*}{Num.} & \multicolumn{2}{c}{Feature} & \multicolumn{3}{c}{Transaction Num.}  \\ 
        \cline{3-7}
                  &   & \multicolumn{1}{c}{code} & \multicolumn{1}{c}{trans.}  & \multicolumn{1}{c}{Min.} & \multicolumn{1}{c}{Max.}  & \multicolumn{1}{c}{$\geqslant$ 100} \\  
        \hline
        Ponzi                         & \textbf{75}     & $\surd$ & $\surd$           & 103      & 105005     & 75 (100\%)                     \\
        Non-Ponzi                     & 325             & $\surd$ & $\surd$           & 100      & 11761    & 325 (100\%)                      \\ 
        \hline
        all                           & 400             & $\surd$ & $\surd$           & 100      &  105005    & 400                      \\
        \hline\hline
        \end{tabular}}
    }{
      \caption{Statastics of the Ethereum Ponzi scheme dataset.}
      \label{tab: dataset}
    }
    \end{floatrow}
    \end{table*}

\subsection{Data Collection} 
Since the known Ponzi schemes published by different platforms are not completely consistent, we collect and collate all current known Ethereum Ponzi scheme data involving contract addresses and corresponding labels from different platforms such as Etherscan\footnote{https://cn.etherscan.com/}, XBlock\footnote{http://xblock.pro/} and Google Cloud\footnote{http://goo.gl/CvdxBp}, as shown in Table~\ref{tab: ponzi}.
Moreover, we consider Ponzi detection as a classification problem, so we select an additional part (325) of normal contract accounts with transaction records as negative samples, and combine them with all Ponzi contract accounts with transaction records (75), yielding an \textbf{Ethereum Ponzi Scheme Dataset (Eth-Ponzi Dataset}, symbolized as $D$ here\textbf{)}, see Table~\ref{tab: dataset}.
Formally, we represent the dataset as $D=\{(a_i, y_i)\}$, where $y_i$ is the identity label reflecting whether the account $a_i$ is a Ponzi account (1 for Ponzi, 0 for non-Ponzi). 

\subsection{Micro Transaction Graph}
The fraudulent behavior of Ponzi schemes is mainly manifested in the interactions between the central Ponzi contract and the surrounding investor accounts.
Hence for a target contract account $a$, we can construct a contract-centric micro transaction graph using all the transaction records related to this account, yielding $g = (a, V,E(t), \textbf{x}^c, \textbf{X}^t,y)$, where node set $V$ consists of accounts involved in these transactions, edge set $E(t)$ consists of related transactions with timestamp.
Note that the node set $V$ consists of the target contract account(CA) $a$, externally owned accounts(EOA) participating in the transactions, and other possible contract acconts.
The CA is controlled by the smart contract code, but the EOA is not, so we use $\textbf{x}^c$ to represent the code features of the target CA $a$, and $\textbf{X}^t$ to represent the transaction features of all accounts.
The code features are 76-dimensional vectors consisting of the frequency of different opcodes, and the transaction features are 15-dimensional manual features that are consistent with \cite{jin2022heterogeneous}.

\subsection{Temporal Evolution Augmentation of Transaction Graph}
As we concern above, existing Ponzi detection methods follow hindsight and utilize the last transaction records (i.e. the full transaction graph) to detect Ponzi schemes, which fails to provide timely warning and sanctions for such financial risks.
In this paper, we propose a \textbf{T}emporal \textbf{E}volution \textbf{Aug}mentation (\emph{TEAug}) strategy for transaction graph with the following purposes:
1) Ponzi schemes on Ethereum begin with the creation and deployment of Ponzi contracts, experience interactions with surrounding investors, and end with the scammer running away with the money or being detected, which is a temporal evolution process.
We use \emph{TEAug} to highlight the micro transaction states of Ponzi contracts in different life cycles.
2) The number of known Ethereum Ponzi schemes is small (several hundred), and direct training with existing data can easily lead to overfitting and low generalization of detection models.
We consider \emph{TEAug} as a data augmentation strategy and increase the scale of trainable data.

\begin{figure}[!t] 
    \centering 
    \includegraphics[width=1\textwidth]{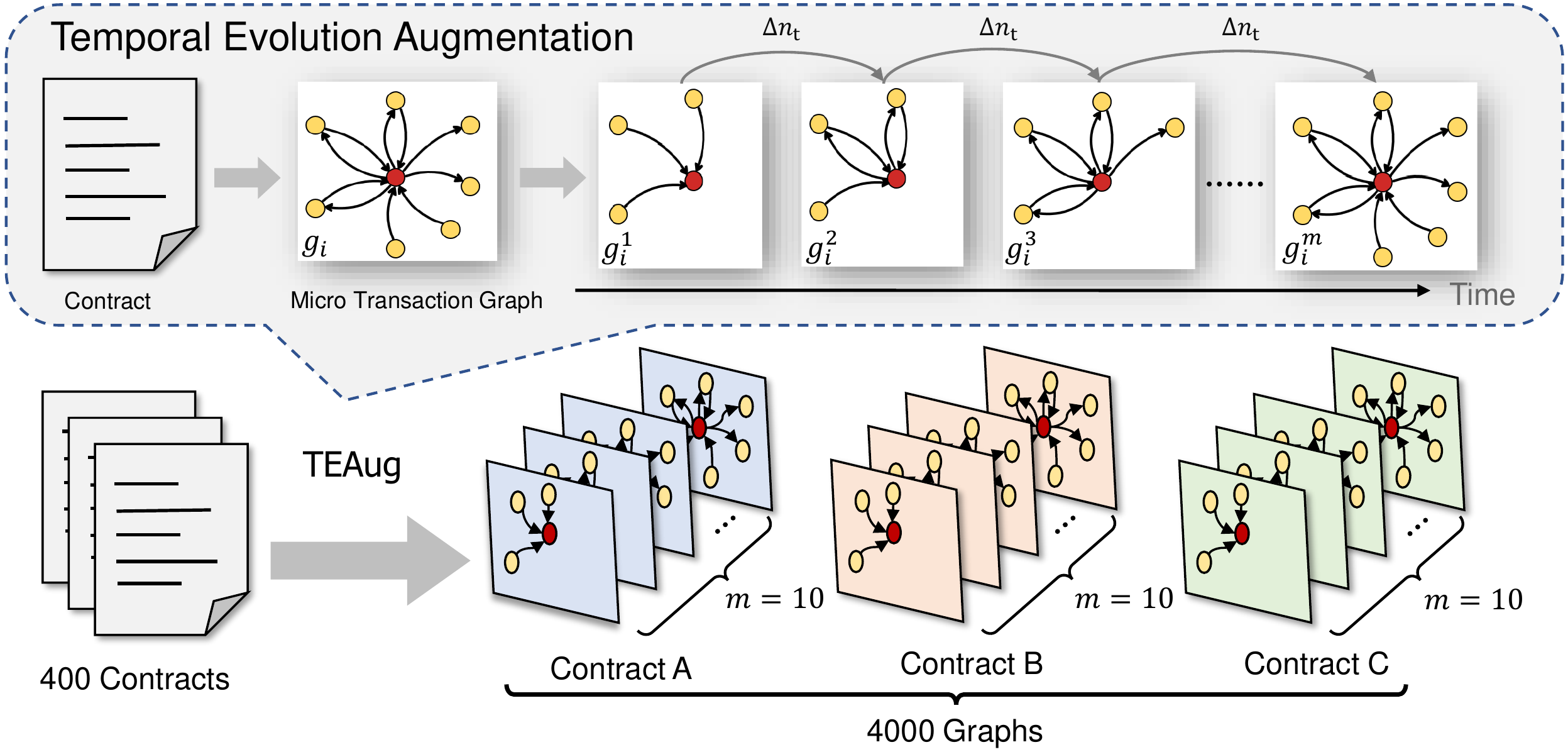} 
    \caption{Illustration of Temporal Evolution Augmentation for Ethereum Ponzi scheme dataset.} 
    \label{fig:TEAug} 
\end{figure}

For each CA in Eth-Ponzi Dataset, we can construct corresponding micro transaction graph, finally yielding a graph set $D_g = \{g_i \mid  \forall a_i \in D\}$.
During \emph{TEAug}, we perform data augmentation on each transaction graph $g_i$ based on a fixed-scale growth of the number of transactions over time (parameterized by $\Delta n_t$), as illustrated in Fig.~\ref{fig:TEAug}.
Formally, for a transaction graph $g_i$, \emph{TEAug} generates a series of transaction subgraphs of different scales based on transaction timestamps:
\begin{equation}
  \begin{aligned}
    \{g_i^1,& g_i^2, \cdots, g_i^m \} \leftarrow  \textsf{TEAug}\left(g_i, \Delta n_t, m\right)   , \\
    & \text{where~~~} g_i^k \subseteq g_i \text{~~and~~}  y(g_i^k) = y(g_i)  \text{~~and~~}  |E_i^{k+1}(t)| =  |E_i^{k}(t)| + \Delta n_t.
  \end{aligned}
\end{equation}
Note that each augmented transaction graph $g_i^k$ is actually the subgraph of $g_i$ and has the same label. 
After \emph{TEAug}, we obtain a larger transaction graph dataset with 4000 samples.

\subsection{Dual-channel Early Warning Model}
\begin{figure}[!t] 
    \centering 
    \includegraphics[width=1\textwidth]{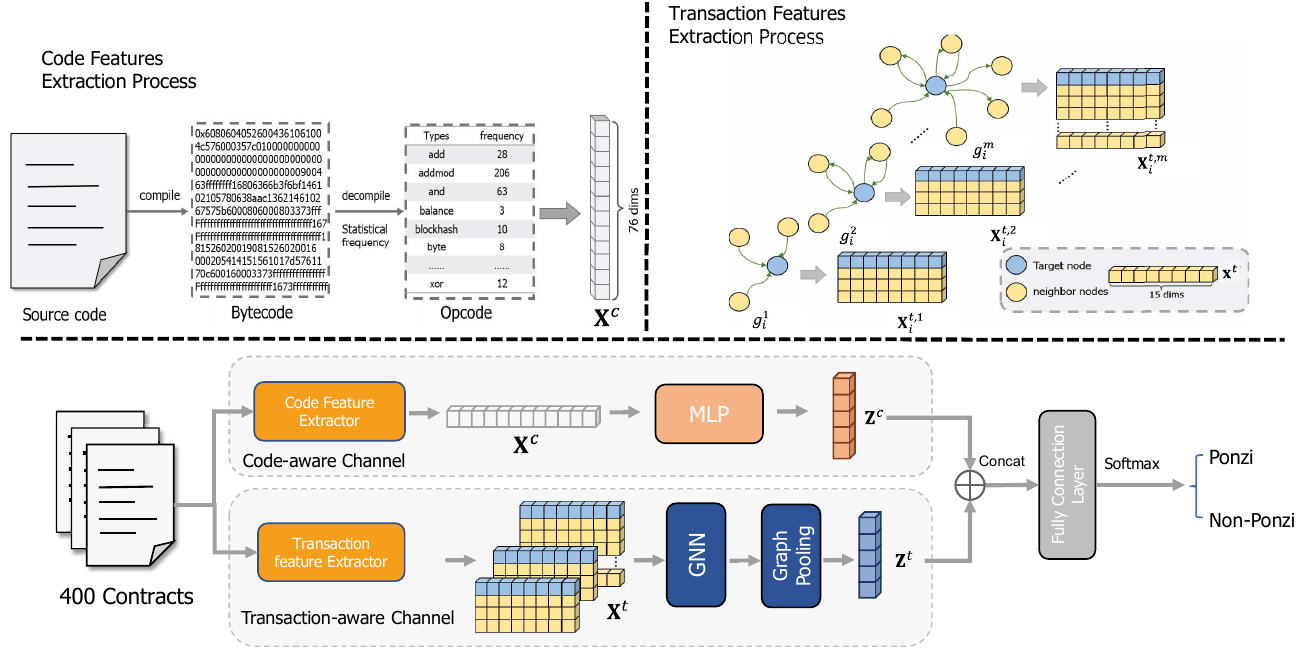}
    \caption{Illustration of dual-channel early warning model for Ethereum Ponzi schemes} 
    \label{fig:framework} 
\end{figure}

In this section, we present a dual-channel early warning model for Ethereum Ponzi schemes, named \emph{Ponzi-Warning}, which focuses on both the contract code and transaction behavior for detecting the occurrence of Ponzi schemes in Ethereum.
As illustrated in Fig.~\ref{fig:framework}, this model mainly consists of two channels: the code-aware channel and the transaction-aware channel.
The former characterizes the code-level features of contract accounts and can take effect at any time after the smart contract is created, and the latter characterizes the transaction behavior patterns of contract accounts and will only take effect after the smart contracts are called and transaction records are generated.

\subsubsection{Code-aware Channel}
For the contract account dataset $D$, we use the feature matrix $\textbf{X}^c \in \mathbb{R}^{n\times f_c}$ to represent the code features of the contract accounts in the dataset, where the $i$-th row of the matrix represents the code feature vector of the $i$-th contract account $a_i$, $n$ is the number of target contract accounts ($n=400$) and $f_c$ is the dimension of initial code features.
In this channel, a multi-layer perceptron (MLP) is applied for capturing the implicit code features:
\begin{equation}
 \textbf{Z}^c =\textsf{MLP}\left(\textbf{X}^c\right).
\end{equation}

\subsubsection{Transaction-aware Channel} 
This channel is a generic feature extraction module that can be compatible with arbitrary GNN layers, aiming at capturing the behavior features implicit in transaction records.
The input of this channel is the lightweight transaction graphs sampled from the augmented dataset and processed through merging multiple edges, and the output is the aggregated graph features $\textbf{Z}^t$.

Formally, graph neural networks (GNN) generalize the convolution operator to irregular graph domains, which can be expressed as a neighborhood aggregation or message passing scheme.
The message pass graph neural networks can be represented as:
\begin{equation}
   \textsf{GNNLayer:~~~} \mathbf{x}_{i}^{(k)}=\gamma^{(k)}\left(\mathbf{x}_{i}^{(k-1)}, \square_{j \in \mathcal{N}(i)} \phi^{(k)}\left(\mathbf{x}_{i}^{(k-1)}, \mathbf{x}_{j}^{(k-1)}\right)\right),
\end{equation}
where $\mathbf{x}_{i}^{(k-1)} \in \mathbb{R}^\textit{F}$ denotes the node features of $v_\textit{i}$ in layer $(k-1)$, $\mathcal{N}(i)$ denotes the neighbor set of $v_\textit{i}$, $\square$ denotes a differentiable aggregation function, e.g., sum, mean or max, and $\gamma$ and $\phi$ denote differentiable functions such as MLPs. 

For an input transaction graph $g$, we stack multiple GNN layers (e.g. two layer) to aggregate and update the node (account) features:
\begin{equation}
      \textbf{H}^t = \textsf{GNNLayer2}\left( \textsf{GNNLayer1}\left(\textbf{X}^t, A\right) \right),
\end{equation}
where $A$ is the adjacency matrix of $g$ that reflects the topological information.
Finally, a graph pooling operation will be performed to obtain the whole-graph representation that reflects the graph-level account behavior pattern features:
\begin{equation}
  \textbf{z}^t = \textsf{GraphPoolingLayer}\left(\textbf{H}^t\right).
\end{equation}

\subsubsection{Dual-channel Joint for Prediction}
The code features output by the code-aware channel and the pattern features output by the transaction-aware channel will be combined through a concatenation operation, and input into a fully connected layer for Ponzi detection:
\begin{equation}
  \textbf{Z} = \textsf{Concat}\left(\textbf{Z}^c, \textbf{Z}^t\right),
\end{equation}
\begin{equation}
  \hat{Y} = \textsf{Softmax}\left(\textsf{MLP}\left(\textbf{Z}\right)\right).
\end{equation}
where $\hat{Y}$ is the set of predicted labels.

\section{Experiments}
\label{sec:4}
In this section, we perform empirical evaluations to demonstrate the effectiveness of the proposed \emph{Ponzi-Warning} framework, answering the following research questions:
\begin{itemize}
  \item [$\bullet$] \textbf{RQ1}: How effective and timely is the proposed \emph{Ponzi-Warning} framework for detecting Ponzi schemes in Ethereum?
  \item [$\bullet$] \textbf{RQ2}: How do code and transaction features affect Ponzi scheme detection?
  \item [$\bullet$] \textbf{RQ3}: What is the appropriate transaction threshold in our framework for reporting Ponzi schemes?
  \item [$\bullet$] \textbf{RQ3}: How does the backbone model affect Ponzi scheme detection?
\end{itemize}

\subsection{Data Setting}
As we can see from Table~\ref{tab: ponzi}, 22.82\% of the Ponzi contracts have no transaction records and 52.01\% of the Ponzi contracts have transaction records of no more than 100.
We filter these contracts with transaction records of less than 100, and use the remaining Ponzi contracts as well as additionally sampled normal contracts, as shown in Table~\ref{tab: dataset}, to evaluate our proposed model.

We split dataset $D$ into training, validation and testing sets with a proportion of $D_\textit{train}:D_\textit{val}:D_\textit{test} = 256:64:80$.
We perform \emph{TEAug} in the three parts:
\begin{equation}
  \begin{aligned}
     &D_\textit{train}^\textit{aug}= \bigcup_{g_i \in D_\textit{train}} \{g_i^1, g_i^2, \cdots, g_i^m \} \leftarrow  \textsf{TEAug}\left(D_\textit{train}, \Delta n_t, m\right)  ,\\
     &D_\textit{val}^\textit{aug}= \bigcup_{g_i \in D_\textit{val}} \{g_i^1, g_i^2, \cdots, g_i^m \} \leftarrow  \textsf{TEAug}\left(D_\textit{val}, \Delta n_t, m\right)  ,\\
     &D_\textit{test}^\textit{aug-i}= \{g_1^i, g_2^i, \cdots, g_{|D_\textit{test}|}^i \}, i\in \{1,2,\cdots, m\} \leftarrow  \textsf{TEAug}\left(D_\textit{test}, \Delta n_t, m\right). \\
  \end{aligned}
\end{equation}
Note that we mix augmented samples of all scales (1 to $m$) to form the augmented training set $D_\textit{train}^\textit{aug}$.
for $D_\textit{val}$, we take augmented samples of all scales to validate our model performance to find the best model parameters.
As for $D_\textit{test}$, we mix all augmented samples of the same scale to form the augmented training set at that scale, finally yielding $m$ augmented validation sets and augmented testing sets, respectively.

\begin{table}[]
  \centering
  \setlength{\tabcolsep}{15pt}
  \renewcommand{\arraystretch}{1}
  \caption{Statastics of augmented datasets (after lightweight) of different scales. \textit{avg.}$|V|$ and \textit{avg.}$|E|$ are the average numbers of nodes and edges, respectively.}
  \begin{tabular}{ccccc} 
    \hline\hline
    \multicolumn{2}{c}{\textbf{Scale~($m=10,\Delta n_t =10$)}} & \multirow{2}{*}{\textit{avg.}$|V|$} & \multirow{2}{*}{\textit{avg.}$|E|$} & \multicolumn{1}{c}{\multirow{2}{*}{\textbf{Proportion}}}  \\ 
    \cline{1-2}
    $i$ & $|E_i^k(t)|$                                         &                                     &                                     & \multicolumn{1}{c}{}                                      \\ 
    \hline
    1    &    10                 & 6.76                 & 6.52                  &        \multirow{10}{*}{75:325}                                  \\
    2    &    20                 & 11.08                & 11.35                 &                                          \\
    3    &    30                 & 15.12                & 15.89                 &                                          \\
    4    &    40                 & 19.11                & 20.34                 &                                          \\
    5    &    50                 & 23.09                & 24.74                 &                                          \\
    6    &    60                 & 27.01                & 29.13                 &                                          \\
    7    &    70                 & 30.89                & 33.50                 &                                          \\
    8    &    80                 & 34.72                & 37.76                 &                                          \\
    9    &    90                 & 38.45                & 41.91                 &                                          \\
    10   &    100                & 42.12                & 46.06                 &                                          \\
    \hline\hline
\end{tabular}
\label{tab:dataset}
\end{table}

\subsection{Baselines}
Manual feature engineering relies on high-performance downstream classifiers to achieve good task performance, so we use three ensemble learning models, XGBoost classifier (XGB), Random forest classifier (RFC) and AdaBoost classifier (ADA), to directly learn the function mapping manual features to account identity.
Meanwhile, since we emphasize the high compatibility of the transaction-aware channel, we use five GNN model for evaluating, namely GCN\cite{kipf2016semi}, GAT\cite{velivckovic2017graph}, ASAPooling\cite{ranjan2020asap}, SAGPooling\cite{lee2019self,knyazev2019understanding} and GlobalAttentionNet(GLAN)\cite{li2015gated}.

\subsection{Experiment Setting}
For the hyperparameter in \emph{TEAug}, we set the scale growth ($\Delta n_t$) as 10, and set the augmentation number ($m$) as 10.
Our model is implemented based on the Pytorch-geometric \footnote{https://github.com/pyg-team/pytorch\_geometric}.
For all methods, the random seed is set as 0, the hidden dimension of each channel is searched from $[16, 32, 64, 128]$, the graph pooling operation is chosen from $[\textsf{MaxPooling}, \textsf{MeanPooling},\textsf{SumPooling}]$, and the transaction-aware channel is implemented using 2-layer GNN. 
In addition, all parameters of the model are initialized using a Gaussian distribution with a mean and standard deviation of 0 and 0.1, respectively. 
And the batch size and L2 penalty are set to 200 and $10^{-5}$, respectively, and a negative log-likelihood loss function is used.
The Adam optimizer is used to optimize these parameters, where the initial learning rate and the dropout rate are set to 0.01 and 0.1, respectively. 

We use $D_\textit{train}^\textit{aug}$ for model training, and validate and test our model using corresponding augmented sets with different scales.
We repeat the experiments five times and report the average F1-score at different stages of the temporal transaction evolution.

\begin{table}
  \centering
  \renewcommand{\arraystretch}{1.5}
  \caption{Performance comparison results w.r.t. F1-score on Ethereum Ponzi scheme dataset.}
  \label{tab: all-reslut}
  \resizebox{\textwidth}{!}{
  \begin{tabular}{ccccccccccccccc} 
  \hline\hline
  \multicolumn{3}{c}{Method}             & \multicolumn{2}{c}{Feature} & \multicolumn{10}{c}{Scale}                                                                                                                                                         \\ 
  \hline
  name          & ~channel-1~ & ~channel-2~  & ~code~    & ~trans.~        & 1               & 2                           & 3               & 4               & 5               & 6                           & 7               & 8                           & 9                           & 10               \\ 
  \hline                                                
  code-aware    & MLP         & $\times$     & $\surd$ &                   & 0.8908          & 0.8908                      & 0.8908          & 0.8908          & 0.8908          & 0.8908                      & 0.8908          & 0.8908                      & 0.8908                      & 0.8908           \\ 
  \hline                                                    
  trans-aware   & $\times$    & GCN          &         & $\surd$           & 0.8283          & 0.8582                      & 0.8685          & 0.8604          & 0.8627          & 0.8713                      & 0.8880          & 0.8858                      & 0.8831                      & 0.8980           \\
  Ours          & no MLP      & GCN          & $\surd$ & $\surd$           & 0.8740          & 0.8794                      & 0.8928          & 0.9056          & 0.9021          & 0.9019                      & 0.9077          & 0.9167                      & 0.9169                      & 0.9164  \\ 
  Ours          & MLP         & GCN          & $\surd$ & $\surd$           & \textbf{0.9054} & \textbf{0.9077}             & \textbf{0.9077} & \textbf{0.9100} & \textbf{0.9100} & \textbf{0.9100}             & \textbf{0.9124} & \underline{\textbf{0.9171}}             & \textbf{0.9171} & \textbf{0.9171}  \\ 
  \hline                            
  trans-aware   & $\times$    & GAT          &         & $\surd$           & 0.8596          & 0.8659                      & 0.8685          & 0.8625          & 0.8627          & 0.8841                      & 0.8892          & 0.8863                      & 0.8938                      & 0.8989           \\
  Ours          & no MLP      & GAT          & $\surd$ & $\surd$           & 0.8802          & 0.8864                      & 0.8962          & 0.8946          & 0.9026          & 0.9017                      & 0.8976          & 0.901                       & 0.9093                      & 0.9093  \\ 
  Ours          & MLP         & GAT          & $\surd$ & $\surd$           & \textbf{0.9012} & \textbf{0.9016}             & \textbf{0.9038} & \textbf{0.9065} & \textbf{0.9134} & \textbf{0.918}              & \textbf{0.9228} & \textbf{0.9228}             & \textbf{0.9202}             & \underline{\textbf{0.9230}}  \\ 
  \hline                                        
  trans-aware   & $\times$    & SAGPooling   &         & $\surd$           & 0.8569          & 0.8888                      & 0.8869          & 0.8894          & 0.9007          & 0.9119                      & 0.9120          & 0.9137                      & 0.9170                      & 0.9154           \\
  Ours          & no MLP      & SAGPooling   & $\surd$ & $\surd$           & 0.8869          & 0.9017                      & 0.9094          & 0.9164          & 0.9164          & 0.9161                      & 0.9244          & 0.9177                      & 0.9232                      & 0.9233  \\ 
  Ours          & MLP         & SAGPooling   & $\surd$ & $\surd$           & \textbf{0.9040} & \textbf{0.9041}             & \textbf{0.9092} & \textbf{0.9212} & \textbf{0.9186} & \textbf{0.9261}             & \textbf{0.9288} & \textbf{0.9288}             & \textbf{0.9288}             & \underline{\textbf{0.9311}}  \\ 
  \hline                                        
  trans-aware   & $\times$    & ASAPooling   &         & $\surd$           & 0.8452          & 0.8783                      & 0.8774          & 0.8915          & 0.9075          & 0.9064                      & 0.9034          & 0.9119                      & 0.9119                      & 0.9119           \\
  Ours          & no MLP      & ASAPooling   & $\surd$ & $\surd$           & 0.8940          & 0.9054                      & 0.9000          & 0.9178          & 0.9199          & 0.9196                      & 0.9199          & 0.9231                      & 0.9231                      & 0.9184  \\ 
  Ours          & MLP         & ASAPooling   & $\surd$ & $\surd$           & \textbf{0.9233} & \textbf{0.9239}             & \textbf{0.9239} & \textbf{0.9239} & \textbf{0.9239} & \underline{\textbf{0.9239}} & \textbf{0.9239} & \textbf{0.9239}             & \textbf{0.9239}             & \underline{\textbf{0.9245}}  \\ 
  \hline                                        
  trans-aware   & $\times$    & GLAN         &         & $\surd$           & 0.8478          & 0.8647                      & 0.8937          & 0.9016          & 0.8908          & 0.9038                      & 0.9109          & 0.9146                      & 0.9017                      & 0.9182           \\
  Ours          & no MLP      & GLAN         & $\surd$ & $\surd$           & 0.8857          & 0.9025                      & 0.8946          & 0.9114          & 0.9009          & 0.9082                      & 0.9154          & 0.9185                      & 0.9186                      & 0.9187  \\ 
  Ours          & MLP         & GLAN         & $\surd$ & $\surd$           & \textbf{0.9164} & \textbf{0.9118}             & \textbf{0.9182} & \textbf{0.9182} & \textbf{0.9182} & \textbf{0.9204}             & \underline{\textbf{0.9280}} & \textbf{0.9280} & \textbf{0.9280}             & \textbf{0.9280}  \\

  \hline\hline
  \end{tabular}}
  \end{table}

\subsection{Evaluation on Ponzi Detection (RQ1)}
To answer \textbf{RQ1}, we evaluate the performance of all the methods in the task of Ponzi scheme detection on Ethereum.
Table~\ref{tab: all-reslut} reports the detection results, from which we can obtain the following \textbf{Obs}ervations.

(1) \emph{\textbf{Obs.1. Both code and transaction features play an important role in Ponzi detection, and our framework achieves the best performance.}}
Overall, our framework outperforms all the other compared frameworks by a significant margin.
When compared with the ``code-aware'' method that only uses code features and MLP model for Ponzi detection, our \emph{Ponzi-Warning} achieves 0.74\% $\sim$ 4.03\% average relative improvement.
When compared with the ``transaction-aware'' method that only uses transaction graph and GNN model for Ponzi detection, our \emph{Ponzi-Warning} achieves 0.98\% $\sim$ 2.41\% average relative improvement.
These phenomena suggest that both the smart contract code and account transaction behavior can expose the information related to account identity.

(2) \emph{\textbf{Obs.2. Our framework enables timely early warning of Ponzi schemes.}}
The transaction graph is constructed based on the transaction records with timestamp information.
During \emph{TEAug}, we only generate the transaction graphs by using transaction records no more than 100, i.e., yielding small-scale transaction graphs.
A smaller-scale transaction graph (with Scale ID $i \leq 10$) shows a less active time of the central contract account.
Our method achieves high and stable detection performance on small-scale transaction graphs, indicating that our \emph{Ponzi-Warning} can achieve timely early warning of Ponzi schemes.

\begin{table}
  \centering
  \renewcommand{\arraystretch}{1.2}
  \caption{The top 5 most important opcode features of Ponzi and non-Ponzi contracts.}
  \label{tab:Code-features}
  \resizebox{\textwidth}{!}{
  \begin{tabular}{c|l|l|c} 
  \hline\hline
  \textbf{Comparison} & \multicolumn{1}{c|}{\textbf{Opcode}} & \multicolumn{1}{c|}{\textbf{Explanation}}                            & \textbf{~~Proportion~~}  \\ 
  \hline
  \multirow{5}{*}{~~Ponzi : non-Ponzi~~} & ~~returndatasize~~   & ~~Size of the last returndata                                            & 207.13       \\
                                         & ~~returndatacopy~~   & ~~Copy the returned data                                                 & 59.22        \\
                                         & ~~codesize      ~~   & ~~Length of the contract                                                 & 6.21         \\
                                         & ~~gaslimit      ~~   & ~~Block gas limit of the current block                                   & 2.53         \\
                                         & ~~smod          ~~   & ~~int256 modulo                                                          & 1.97         \\ 
  \hline
  \multirow{5}{*}{~~non-Ponzi : Ponzi~~} & ~~delegatecall  ~~   & ~~Calling another contract's method using the current contract's store~~ & 17.31      \\
                                         & ~~create2       ~~   & ~~Create a subcontract using a defined address                           & 14.31      \\
                                         & ~~byte          ~~   & ~~Return (u)int256 x the i-th byte starting from the highest byte        & 5.31       \\
                                         & ~~difficulty    ~~   & ~~Difficulty of the current block                                        & 4.62       \\
                                         & ~~sar           ~~   & ~~int256 right shift                                                     & 3.92       \\
  \hline\hline
  \end{tabular}}
\end{table}

\subsection{Single Channel Analysis (RQ2)}
To answer \textbf{RQ2}, we analyze the performance of each channel independently and obtain the following \textbf{Obs}ervations.

(1) \emph{\textbf{Obs.3. The characteristics of the code carry the purpose of the contract creation.}}
Code features accompany the entire life cycle of a contract account, and we can use code features to perform Ponzi detection at the initial stage of contract creation.
From Table~\ref{tab: all-reslut}, we can observe that using only code features has achieved detection accuracy up to 89\%, far surpassing the method of using transaction features in the early stages, indicating that there is a strong correlation between the characteristics of the contract code and the legitimacy of the contract account.
We list the top 5 most important opcodes of Ponzi and non-Ponzi contracts according to the feature importance analysis of RFC, as shown in Table~\ref{tab:Code-features}.
We observe that Ponzi contracts involve more ``returndata'' opcodes and will make plenty of judgments about the ``returndata''.
On the contrary, non-Ponzi contracts involve more ``delegatecall'' and ``create'' opcodes, which suggests that normal contracts would have a large number of methods to call other contracts, while the nature of Ponzi schemes is to attempt to carry out as much fraud as possible in one contract.

\begin{figure}
  \centering 
  \includegraphics[width=1\textwidth]{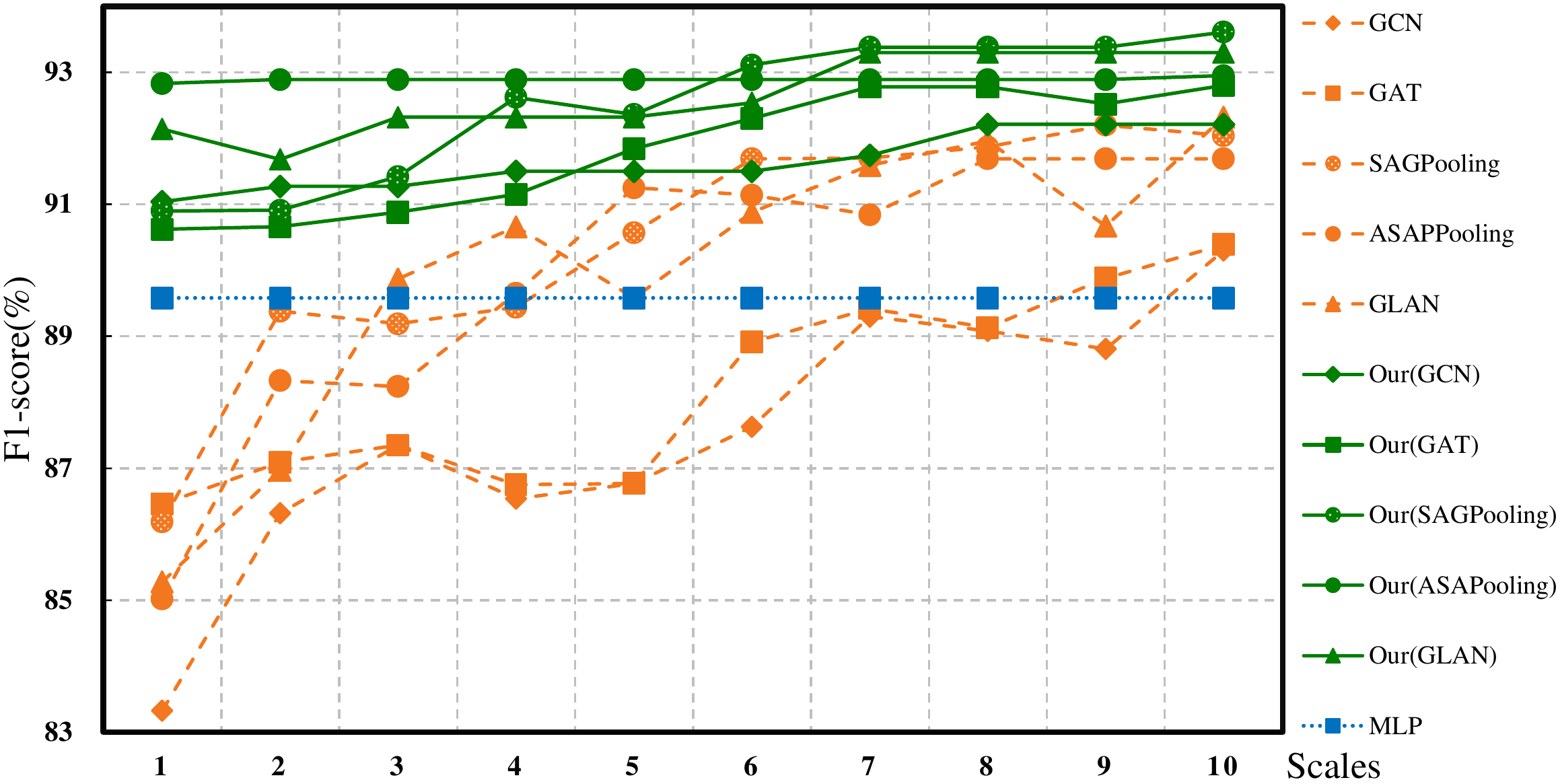} 
  \caption{Performance comparison between our framework and the ordinary graph neural network methods under different scales of transactions.} 
  \label{pic:all_reslut_pic} 
\end{figure}

(2) \emph{\textbf{Obs.4. The characteristics of the transaction records expose the purpose of the contract creation over time.}}
Transaction features can only be available after transaction has occurred.
From Table~\ref{pic:all_reslut_pic}, we can observe that with few transaction records (with Scale ID $i = 1$), the transaction-aware channel achieves poor detection performance.
And the performance of this channel is greatly improved as the size of the transaction graph increases, indicating that more transaction records can expose more account purpose, thereby improving the Ponzi detection.
Furthermore, we find that transaction-aware channel can achieve a higher performance limit than code-aware channel, which inspires us to perform dynamic detection for better early warning.

\subsection{Threshold of Reporting Ponzi Schemes (RQ3)}
To answer \textbf{RQ3}, we further analyze the performance curve in Fig.~\ref{pic:all_reslut_pic} in terms of detection accuracy and timeliness.
The two types of curves (GNNs and our framework) all show a certain upward trend, in which the former shows a poor initial performance and a large increase, while the latter shows a power initial performance and a stable and small increase.
In our framework, the high initial performance benefits from the blessing of the code channel, and the subsequent increase benefits from the temporal evolution of the transaction.
We further observe that in most cases, the performance curves rise first, reach the critical point when the Scale ID equals to 7, and then gradually stabilize.
So we reasonably consider 70 (with Scale ID $i = 7$) as the appropriate transaction threshold for reporting Ponzi schemes.

\begin{figure}
  \centering 
  \includegraphics[width=1\textwidth]{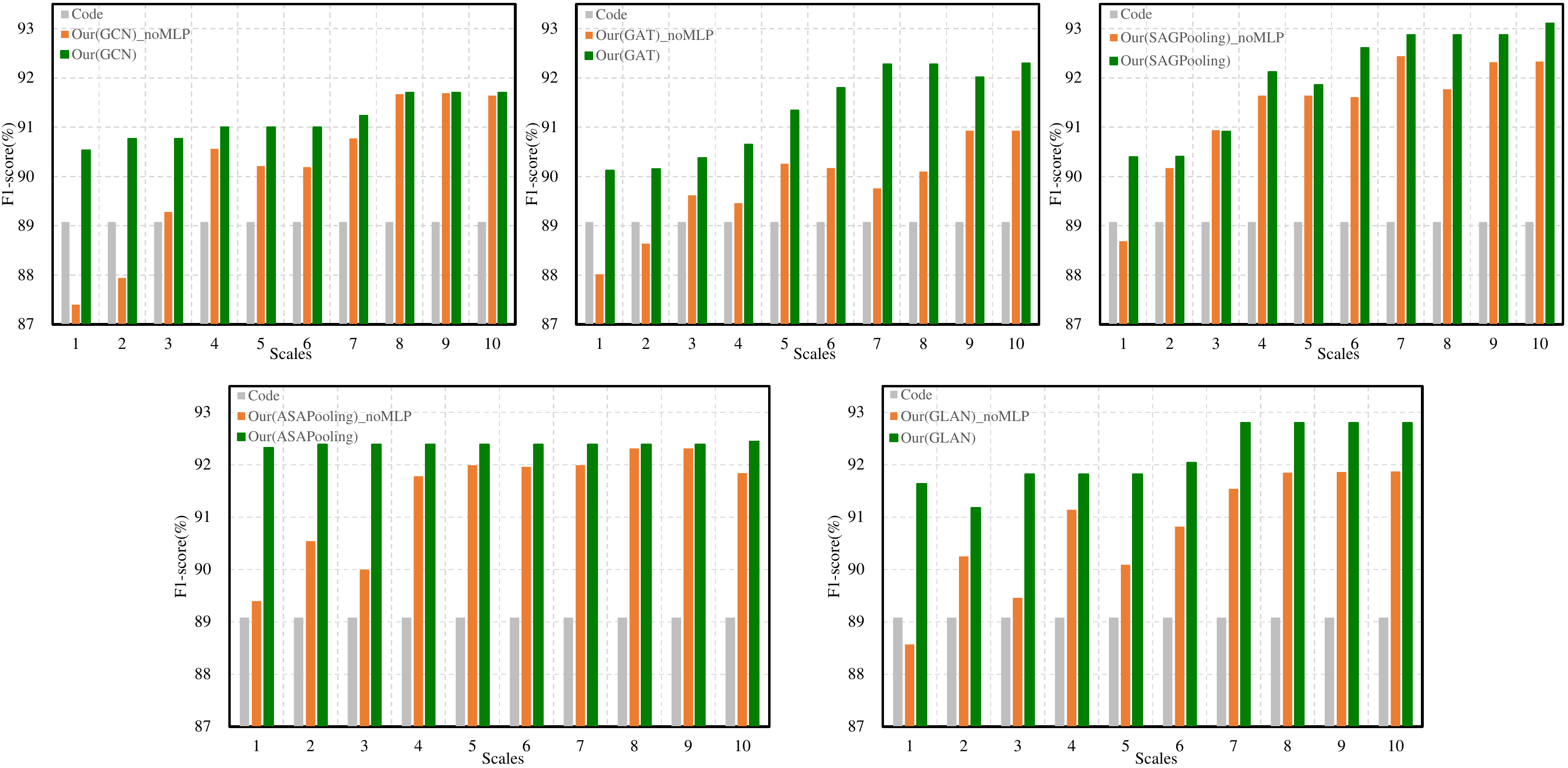} 
  \caption{Performance comparison between code-aware channels with and without MLP.} 
  \label{pic:five_GNN_no_MLP} 
\end{figure}

\subsection{Ablation Study}
To answer \textbf{RQ4}, we perform an ablation study by removing or replacing the backbone models in the two channels.
As shown in Fig.~\ref{pic:five_GNN_no_MLP} and \ref{pic:only_trans_M_G_pic}, the corresponding observation results have the following aspects:

(1) \emph{\textbf{Obs.5. Further feature extraction on initial opcode features benefits Ponzi detection.}}
We first remove the backbone model (MLP) in the code-aware channel, and directly concatenate the unprocessed initial code features and the output of the transaction-aware channel, yielding the comparison graph \textsf{Our(GNN)\_noMLP}.
As shown in Fig.~\ref{pic:five_GNN_no_MLP}, we observe that \textsf{Our(GNN)} always outperforms \textsf{Our(GNN)\_noMLP} across all scales, suggesting that implicit information extracted from initial code features can significantly improve the Ponzi detection.
This phenomenon encourages us to further design more powerful backbone models in the code-aware channel.

(2) \emph{\textbf{Obs.6. Powerful GNN models can capture the account interaction behavior benefitting Ponzi detection.}}
We replace the backbone model (GNN) in the transaction-aware channel with machine learning models, and compare the performance difference.
As shown in Fig.~\ref{pic:only_trans_M_G_pic}, we observe that three more powerful GNN models outperform machine learning methods while GCN and GAT obtain relatively poor performance, indicating that the structural interaction behavior features captured by the GNN models can improve the Ponzi detection.

\section{Conclusion}
\label{sec:5}
In this paper, we propose a dual-channel early warning framework for Ethereum Ponzi schemes, which can effectively extract and fuse code-level and transaction-level features.
Experimental results demonstrate that our framework can achieve effective and timely early warning for Ponzi schemes.
Moreover, we also analyze the code and transaction level differences between normal and Ponzi accounts.
In the future work, we will consider designing a more powerful code-aware channel and replacing the transaction-aware channel with a dynamic graph neural network model for better Ponzi detection.

\subsubsection*{Acknowledgments}
This work was partially supported by the National Key R\&D Program of China under Grant 2020YFB1006104, by the Key R\&D Programs of Zhejiang under Grants 2022C01018 and 2021C01117, by the National Natural Science Foundation of China under Grant 61973273 and 62103374, and by the Zhejiang Provincial Natural Science Foundation of China under Grant LR19F030001, and by Basic Public Welfare Research Project of Zhejiang Province Grant LGF20F020016 and Open Project of the Key Laboratory of Public Security Informatization Application Based on Big Data Architecture Grant 2020DSJSYS003.

\bibliographystyle{splncs04_}
\bibliography{ref}

\end{document}